\documentclass{elsart}

\usepackage{graphics}
 \usepackage{graphicx}
 \usepackage{epsfig}

\usepackage{amssymb}
\begin{document}
\hfill WUE-ITP-2000-023\\
\hspace*{\fill} hep-ph/xxxxxxx\\

\begin{frontmatter}


\title{Distinguishing between MSSM and NMSSM with $e\gamma$ Scattering \thanksref{dfg}
}
\thanks[dfg]{Work supported by the Deutsche Forschungsgemeinschaft, contract FR 1064/4-1 and the Bundesministerium f\"ur Bildung und Forschung, contract 05 HT9WWA 9.}
\author{Claus Bl\"ochinger\thanksref{label1}},
\author{Fabian Franke\thanksref{label2}}, 
\author{Hans Fraas\thanksref{label3}}
\thanks[label1]{e-mail:bloechi@physik.uni-wuerzburg.de}
\thanks[label2]{e-mail:fabian@physik.uni-wuerzburg.de}
\thanks[label3]{e-mail:fraas@physik.uni-wuerzburg.de}

\address{Institut f\"ur Theoretische Physik, Universit\"at W\"urzburg, Am Hubland, D-97074 W\"urzburg, Germany
}


\begin{abstract}
We study the associated production of selectrons and neutralinos with 
subsequent leptonic decay of the selectron in 
$e\gamma$ scattering within the framework of MSSM and NMSSM.
Due to the weak couplings of singlino dominated neutralinos, associated neutralino
production
$e^+e^- \rightarrow\tilde{\chi}_1^0\tilde{\chi}_2^0$ at a linear collider can be strongly suppressed in
some regions of the parameter space. Then the process 
$e\gamma\rightarrow \tilde{e}_{L/R}\tilde{\chi}_{1/2}^0\rightarrow e\tilde{\chi}_{1/2}^0\tilde{\chi}_1^0$ can give additional informations
on the underlying theory and the character of $\tilde{\chi}_1^0$ and 
$ \tilde{\chi}_2^0$.


\vspace*{0.3cm}
\hspace{-0.4cm}{\sl PACS:} 11.30.Pb, 12.60.Jv, 14.80.Ly
\end{abstract}


\end{frontmatter}

\section{Introduction}

Supersymmetry is up to now one of the most attractive theories beyond the 
Standardmodel (SM). But from the theoretical point of view there exist 
many ways
for a supersymmetric extension of the SM. One of the widely used models
is the
Minimal Supersymmetric Standard Model (MSSM) that has minimal content in the 
Higgs sector. The masses, mixings and couplings of the four neutralinos are
determined by the parameters 
$M_1$, $M_2$, $\mu$ and $\tan\beta$.
The minimal extension of the MSSM by a higgs singlet field $S$ is the 
Next-to-Minimal
Supersymmetric Standard Model (NMSSM) \cite{nmssm1}. The neutralino sector now contains five
neutralinos being mixtures of the MSSM gauginos and higgsinos and an additional 
singlino $\tilde{S}$. The NMSSM neutralino mixing depends on the parameters
$M_1$, $M_2$, $\tan\beta$, the singlet vacuum expectation value $x$ and the trilinear
couplings $\lambda$ and $\kappa$ \cite{nmssmneu}. For large values of $x$ the singlino
dominated neutralino decouples and the remaining four neutralinos have
the same masses and characters as in the MSSM if one identifies $\mu=\lambda x$ \cite{stefan1}.

 The linear collider (LC) will first run in the $e^+e^-$-mode.
However, associated neutralino production $e^+e^- \rightarrow \tilde{\chi}_1^0
\tilde{\chi}_2^0$ may not give sufficient evidence for a discrimination between
MSSM and NMSSM, especially if this is the only kinematical accesible neutralino
production process in the $e^+e^-$-mode. In our contribution we want to investigate the possibilities to distinguish
between the models with the help of $e\gamma$ scattering $e\gamma\rightarrow \tilde{e}_{L/R}\tilde{\chi}_{1/2}^0\rightarrow e\tilde{\chi}_{1/2}^0\tilde{\chi}_1^0$.

\section{Cross sections of $e\gamma\rightarrow \tilde{e}_{L/R}\tilde{\chi}_{1}^0\rightarrow e\tilde{\chi}_{1}^0\tilde{\chi}_1^0$}

We first study the associated production of left/right selectrons and the lightest neutralino
(LSP) 
in $e\gamma$ scattering. The high energetic photon comes from 
Compton back-scattering of laser photons off one electron beam at a LC.  The production proceeds via
s-channel exchange of an electron and t-channel exchange of selectrons.
The formulas for special cases are given in \cite{egamma1}, the complete
analytical expressions for the differential and total cross sections for
polarized beams will be given in a forthcoming paper \cite{egamma2}. 
The 
subsequent decay of the selectrons in electrons and the LSP leads to an $e^-$
and missing energy in the final state.
The total cross section for the production and
decay in the $ee$-cms depends on 
the polarisation $P_e$ of the electron beam, the helicity
$\lambda_L$ of the laser photon and the helicity $\lambda_e$ of the converted
electron beam. 

In fig.~1 we show the contours of constant total cross sections for the process
$e^-\gamma\rightarrow \tilde{\chi}_1^0\tilde{e}_{L/R}\rightarrow e^-\tilde{\chi}_1^0\tilde{\chi}_1^0$ in the 
$M_2$-$\mu$-plane for the MSSM and the $M_2$-$x$-plane for the NMSSM for 
$\sqrt{s_{ee}}=500$ GeV. 
For the calculations we choose the longitudinal polarisation of the electron 
beam $P_e=80\%$, the helicity of the converted electron beam $\lambda_e=-80\%$
and the helicity of the laser photon $\lambda_L=100\%$. This set of 
polarisations can be achieved at TESLA and leads to the most significant 
contours. We set $\tan\beta=3$, 
the common scalar mass 
$m_0=110$ GeV and use the GUT relation $M_1=5/3\tan^2\theta_W M_2$. 
The trilinear couplings are choosen as $\lambda=0.5$ and 
$\kappa=0.1$ (fig.~1b), $\lambda=0.5$ and $\kappa=0.01$ (fig.~1c) and
$\lambda=0.1$ and $\kappa=0.01$ (fig.~1d). 

If in the NMSSM the $\tilde{\chi}_1^0$ is dominated by the singlino content the production is
supressed, because electron and selectron only couple to the
gaugino components of the neutralino. Then  the total cross section is smaller 
than 1 fb in some parameter regions (fig.~1b-d) and probably lies below the 
detection limit of a $e\gamma$ collider. Otherwise one obtains  cross sections similar to the MSSM.
 For $M_2>0$ there also exist a small
band with 1 fb $<\sigma<$ 10 fb (fig.~1c/d). In this region the
$\tilde{\chi}_1^0$ is the mainly singlino-like, but 
nevertheless it will be very difficult to extract this small signal from 
the background $e\gamma\rightarrow W\nu$ \cite{background}.

\begin{figure}[htb]
\label{fig1}
\centering
\begin{picture}(13.9,12.8)
\put(-1.2,1.9){\includegraphics{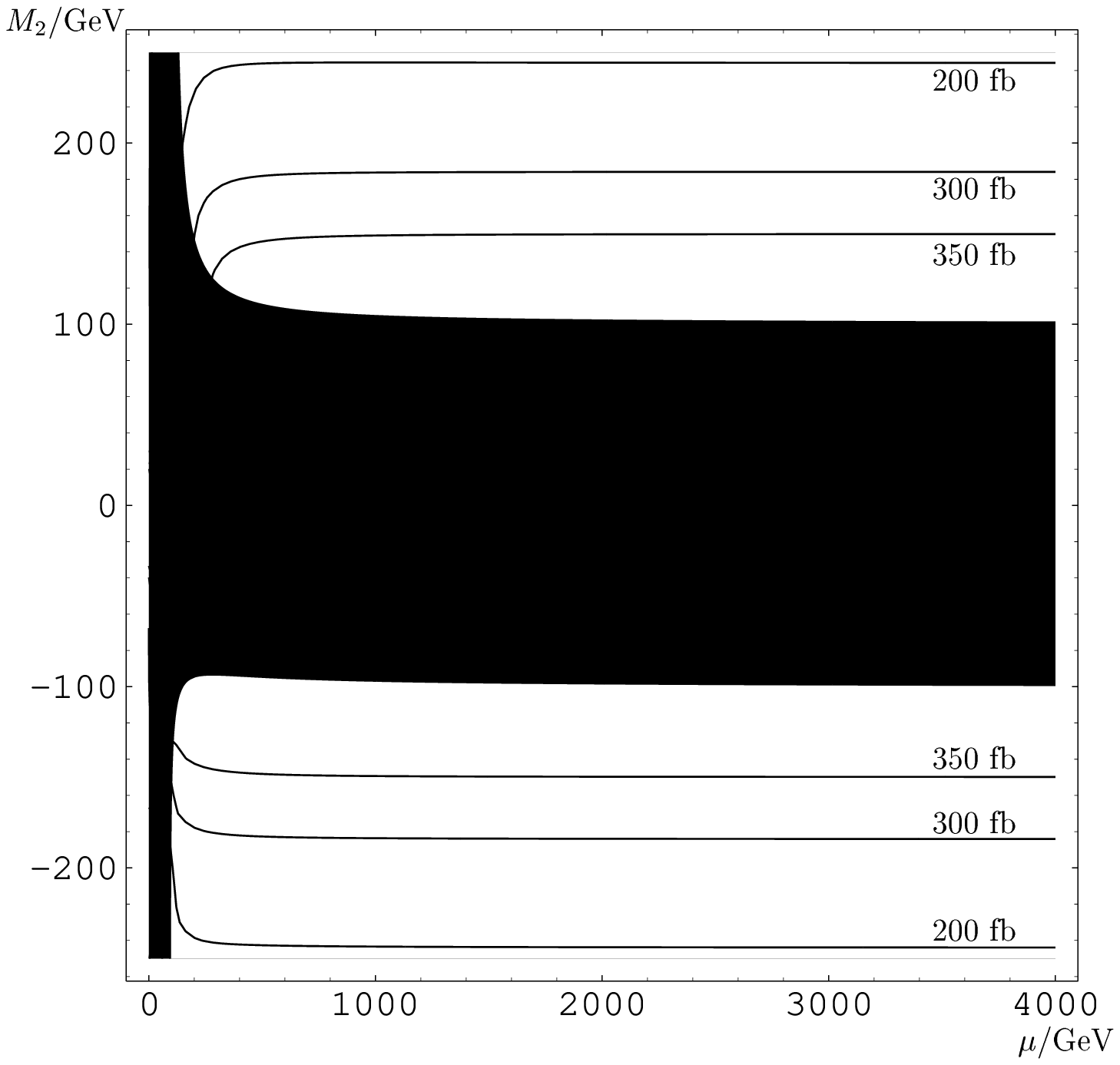}}
\put(3.2,12.6){{\tiny (a)}}
\put(3.2,5.9){{\tiny (c)}}
\put(10.38,12.6){{\tiny (b)}}
\put(10.38,5.9){{\tiny (d)}}
\put(6.0,1.9){\includegraphics{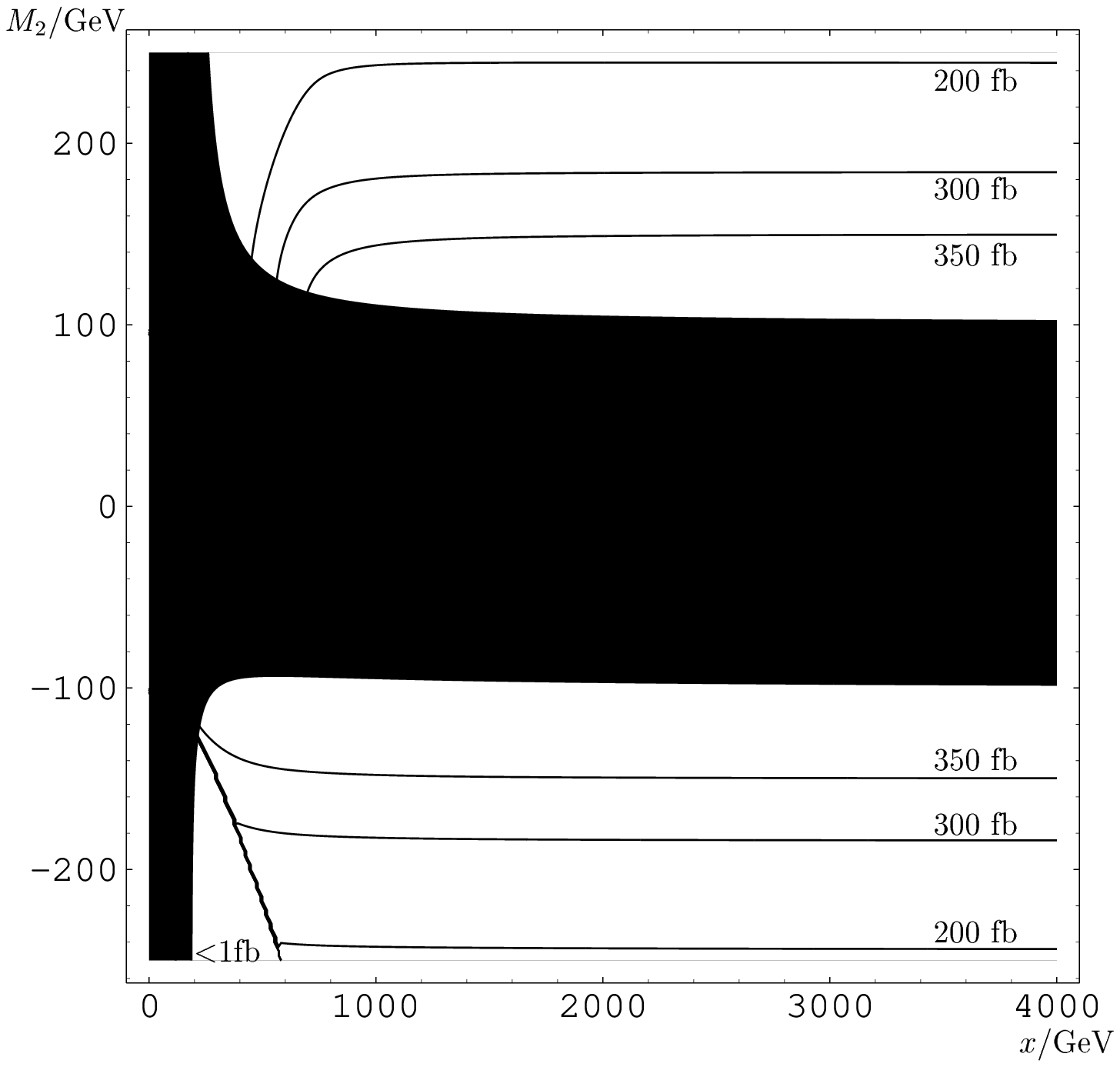}}
\put(-1.2,-4.8){\includegraphics{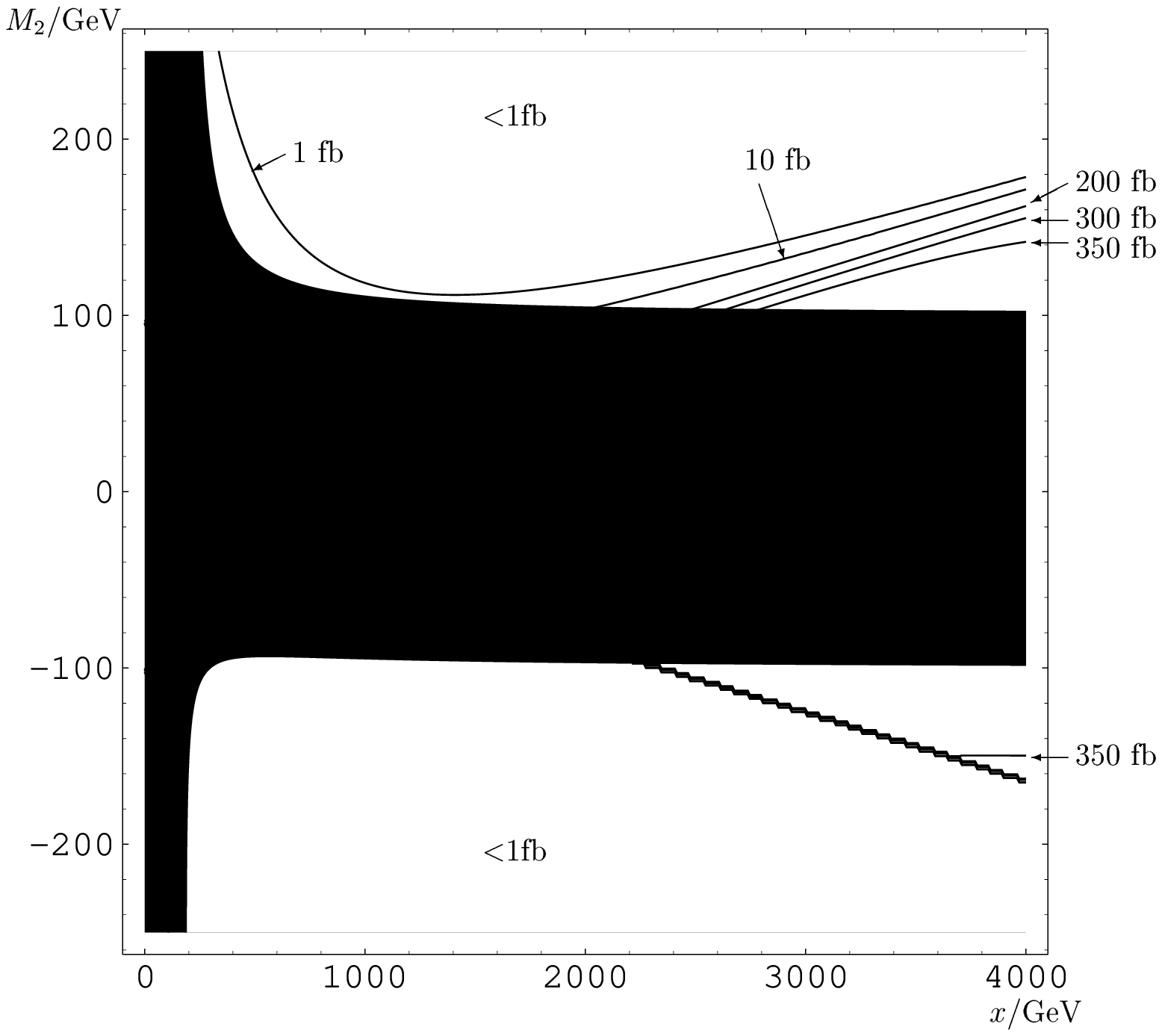}}
\put(6.0,-4.8){\includegraphics{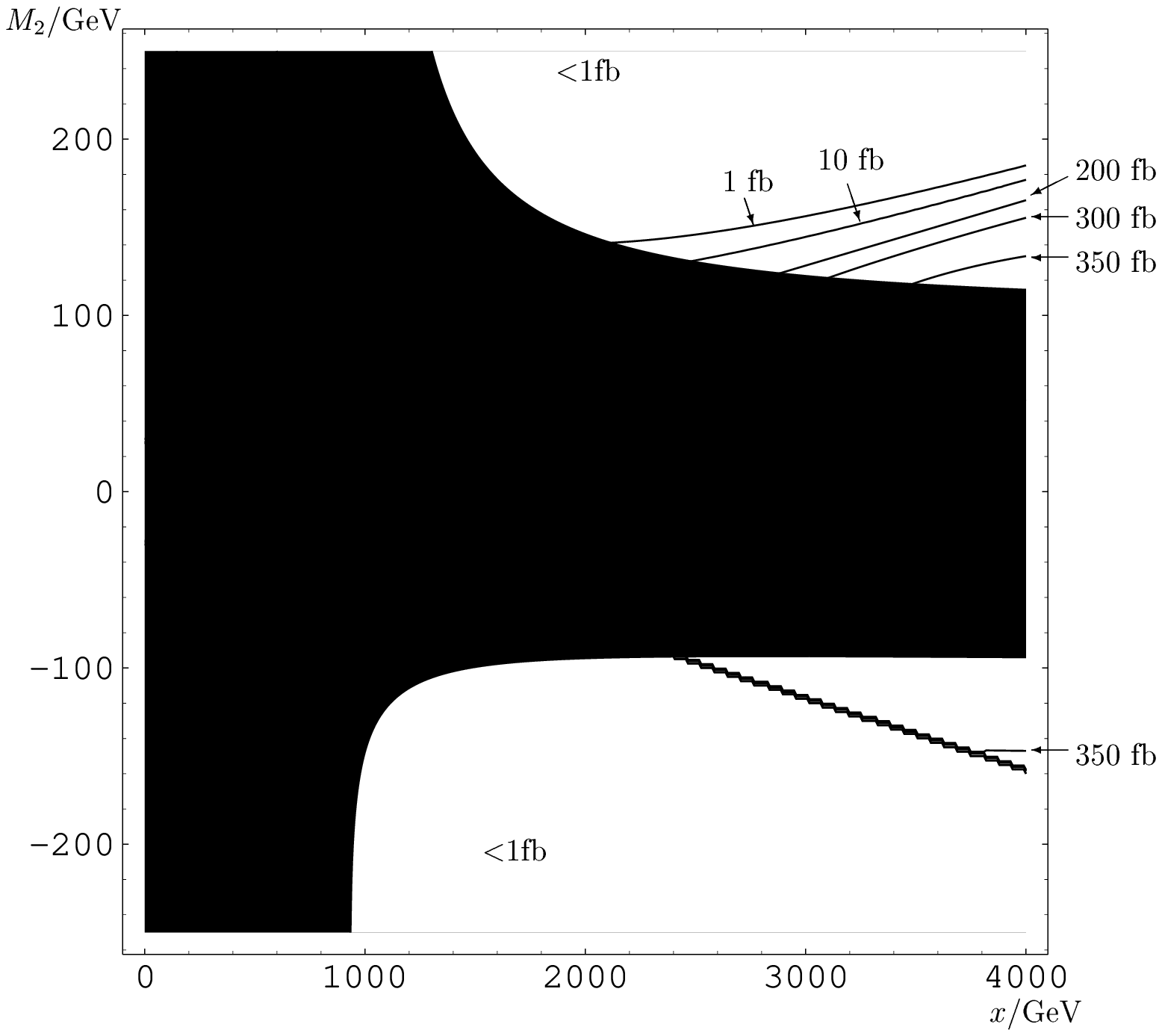}}
\end{picture}
\caption{Contours of constant total cross sections for the process 
$e^-\gamma\rightarrow \tilde{\chi}_1^0\tilde{e}_{L/R}\rightarrow e^-\tilde{\chi}_1^0\tilde{\chi}_1^0$ for $\sqrt{s_{ee}}=500$ GeV, polarisations 
$P_e=80\%$ of the electron beam, $\lambda_e=80\%$ of the converted 
electron beam, $\lambda_L=100\%$ of the laser photon and the parameters 
$\tan\beta=3$, $m_0=110$ GeV : 
(a) MSSM; (b) NMSSM with $\lambda=0.5$ and
$\kappa=0.1$; (c) NMSSM with $\lambda=0.5$ and
$\kappa=0.01$; (d) NMSSM with $\lambda=0.1$ and
$\kappa=0.01$.} 
\end{figure}

\section{Distinguishing NMSSM and MSSM with $e\gamma$ Scattering}

In this section we consider two cases: 
(a) $\tilde{\chi}_1^0$ has singlino character and  
$\tilde{\chi}_2^{0, NMSSM}$ corresponds to
$\tilde{\chi}_1^{0,MSSM}$; (b) $\tilde{\chi}_2^0$ is singlino dominated
and $\tilde{\chi}_1^0$ is similar in NMSSM and MSSM. 
In table 1 we present four reference scenarios I, II (a) and III, IV (b) with fixed $m_{\tilde{\chi}_1^0}$ 
and $m_{\tilde{\chi}_2^0}$ for comparison. In all scenarios the lightest MSSM-like
neutralino is dominated by gaugino components, which couple to electrons
and selectrons, while the higgsino components do not. 
For $m_{\tilde{\chi}_2^0}>\sqrt{s_{ee}}/2$ and
$m_{\tilde{\chi}_1^0}+m_{\tilde{\chi}_3^0}>\sqrt{s_{ee}}$, the only channel
for direct production
of neutralinos in $e^+e^-$ annihilation is $e^+e^-\rightarrow \tilde{\chi}_1^0
\tilde{\chi}_2^0$. However, if one of the neutralinos is mainly singlino this
process will be suppressed.
The cross sections for the process $e^+e^-\rightarrow \tilde{\chi}_1^0
\tilde{\chi}_2^0$ in fig.~2a are calculated with longitudinal polarisations 
$P_{e^-}=80\%$ of the electron beam and 
$P_{e^+}=-60\%$ of the positron beam.
Fig.~2b/c shows the cross sections for $e^-\gamma\rightarrow \tilde{\chi}_{1/2}^0\tilde{e}_{R}\rightarrow e^-\tilde{\chi}_{1/2}^0\tilde{\chi}_1^0$ with
polarisations $P_e=80\%$,
$\lambda_e=-80\%$ and $\lambda_L=100\%$ in the reference scenarios.
In all scenarios we fix the masses $m_{\tilde{e}_R}=180$ GeV and $m_{\tilde{e}_L}=300$ GeV.

\begin{center}
\begin{table}
\centering
\begin{tabular}{|c|c|c|c|c|}
\hline
& {\bf Scenario I}& {\bf Scenario II}& {\bf Scenario III}& {\bf Scenario IV}\\
\hline
\hline
$M_2$/GeV & 520. & -513.9 & 353.0 & -346.3\\
\hline 
$x$/GeV & \multicolumn{4}{c|}{1600.}\\
\hline
$\lambda$ & \multicolumn{4}{c|}{0.5}\\
\hline
$\kappa$ & 0.0531 & 0.05295 & 0.0794 & 0.07923\\
\hline
$m_{\tilde{\chi}_1^0}$/GeV & (-)173.4 & (-)173.4 & (-)173.4 & (+)173.4 \\
\hline
$m_{\tilde{\chi}_2^0}$/GeV & (-)256.6 & (+)256.6 & (-)256.6 & (-)256.6 \\
\hline
$m_{\tilde{\chi}_3^0}$/GeV & (-)504.3 & (+)513.5 & (-)343.8 & (+)348.1 \\
\hline
$\langle\tilde{S}\vert\tilde{\chi}_1^0\rangle^2$ & 0.98 & 0.99&0.0024 &0.0008\\
\hline
$\langle\tilde{S}\vert\tilde{\chi}_2^0\rangle^2$ & 0.0027 &0.0009&0.979 &0.99\\
\hline
\end{tabular}
\caption{Reference scenarios with fixed $m_{\tilde{\chi}_1^0}$ and 
$m_{\tilde{\chi}_2^0}$ (signs of the mass eigenvalues in brackets). In scenario I and II the $\tilde{\chi}_1^0$ is mainly 
singlino and in scenario III and  IV $\tilde{\chi}_2^0$ is 
singlino-like.}
\end{table}
\end{center}

\subsection{If  $\tilde{\chi}_1^0$ is singlino-like...}

... we expect for $e^+e^-\rightarrow \tilde{\chi}_1^0\tilde{\chi}_2^0$ a  
production cross section of 
$\sigma=0.9$ fb for scenario I and $\sigma=0.27$ fb for scenario II at 
$\sqrt{s_{ee}}=500$ GeV (fig.~2a). 
The smaller cross section in scenario II is due to the different
relative signs of $m_{\tilde{\chi}_1^0}$ and $m_{\tilde{\chi}_2^0}$, which
leads to cancellations in the interference terms 
\cite{fabian}. Now it will be a question of luminosity and background
supression, whether the cross sections are measurable.
Even if a signal is detected, the decision whether $\tilde{\chi}_1^0$
or $\tilde{\chi}_2^0$ is the singlino dominated neutralino will be difficult 
since the cross sections are very similar ($\sigma=0.9$ fb in scenario I and 
$\sigma=1.1$ fb in scenario III, see fig.~2a). 
In $e\gamma$ scattering the process 
$e^-\gamma\rightarrow \tilde{\chi}_1^0\tilde{e}_{R}\rightarrow e^-\tilde{\chi}_1^0\tilde{\chi}_1^0$ will be strongly supressed (see sec.~1). Then the  first 
detectable 
neutralino production process at a high luminosity collider will be 
$e^-\gamma\rightarrow \tilde{\chi}_2^0\tilde{e}_{R}\rightarrow e^-\tilde{\chi}_2^0\tilde{\chi}_1^0$. Fig.~2c shows  
total cross sections of $\sigma=3.8$ fb in scenarios I and II at 
$\sqrt{s_{ee}}=500$ GeV which further increase for higher energies.
This process, however, will lead to a completely different 
final state compared to the LSP-selectron production because of the additional decay 
 of the $\tilde{\chi}_2^0$ \cite{fabian}. Therefore one can decide via
detection of the $e\gamma$ process that the $\tilde{\chi}_1^0$ is mainly
 singlino.

If the $e^+e^-$ cross section is too small for detection, the production and
decay of $\tilde{\chi}_2^0$ together with $\tilde{e}_{R}$ in $e\gamma$ will be the 
first 
direct production process of neutralinos that would be measurable at
$\sqrt{s_{ee}}=500$ GeV with high luminosity. The non-detection
of the $\tilde{\chi}_1^0\tilde{e}_{R}$ production suggests a singlino-like
$\tilde{\chi}_1^0$ and therefore extended supersymmetry  as the 
underlying theory.

\begin{figure}[htb]
\label{fig2}
\centering
\begin{picture}(13.9,8.8)
\put(3.2,8.6){{\tiny (a)}}
\put(10.38,8.6){{\tiny (b)}}
\put(6.8,4.2){{\tiny (c)}}
\put(-1.2,-2.0){\includegraphics{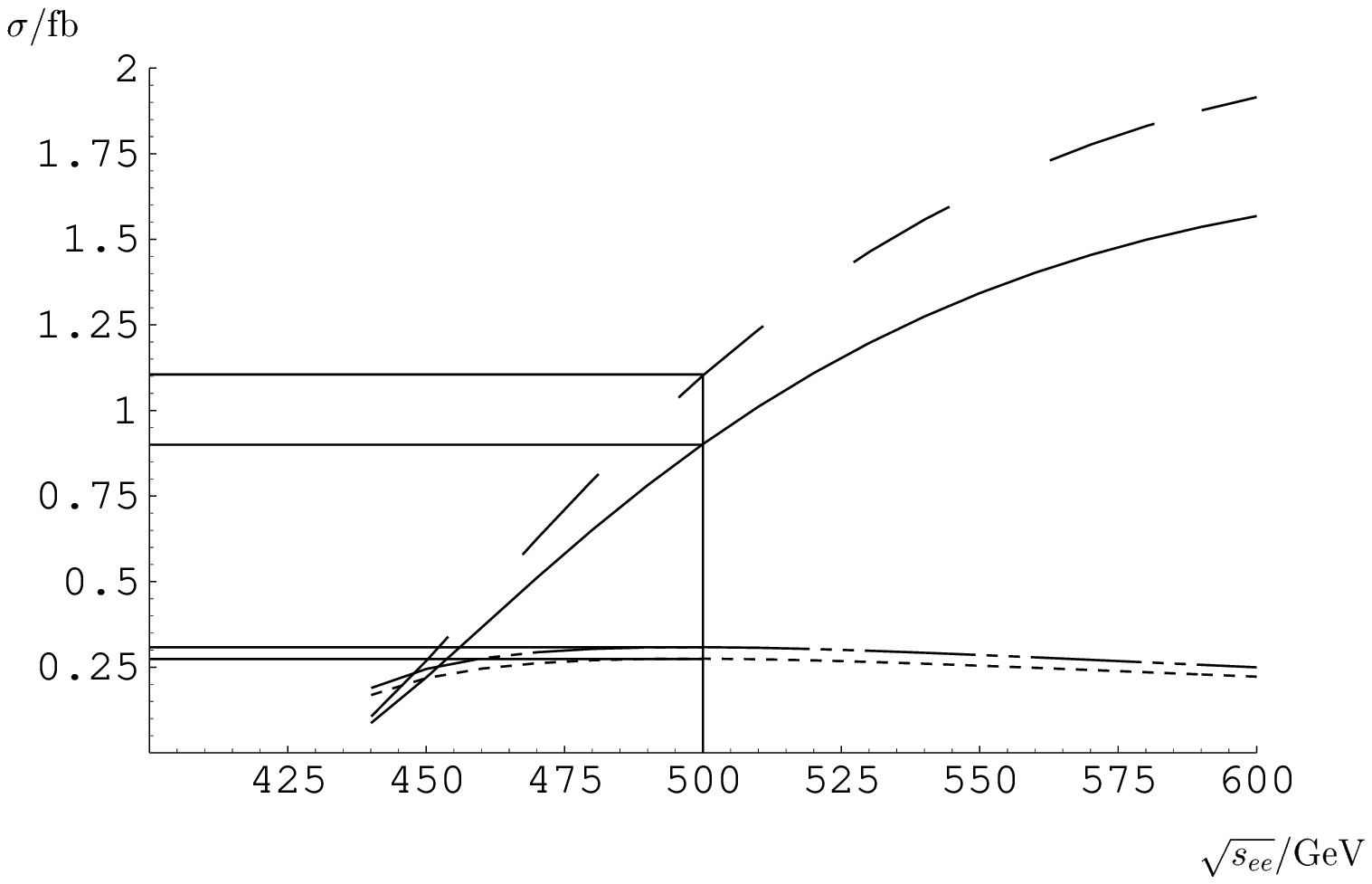}}
\put(6.0,-2.0){\includegraphics{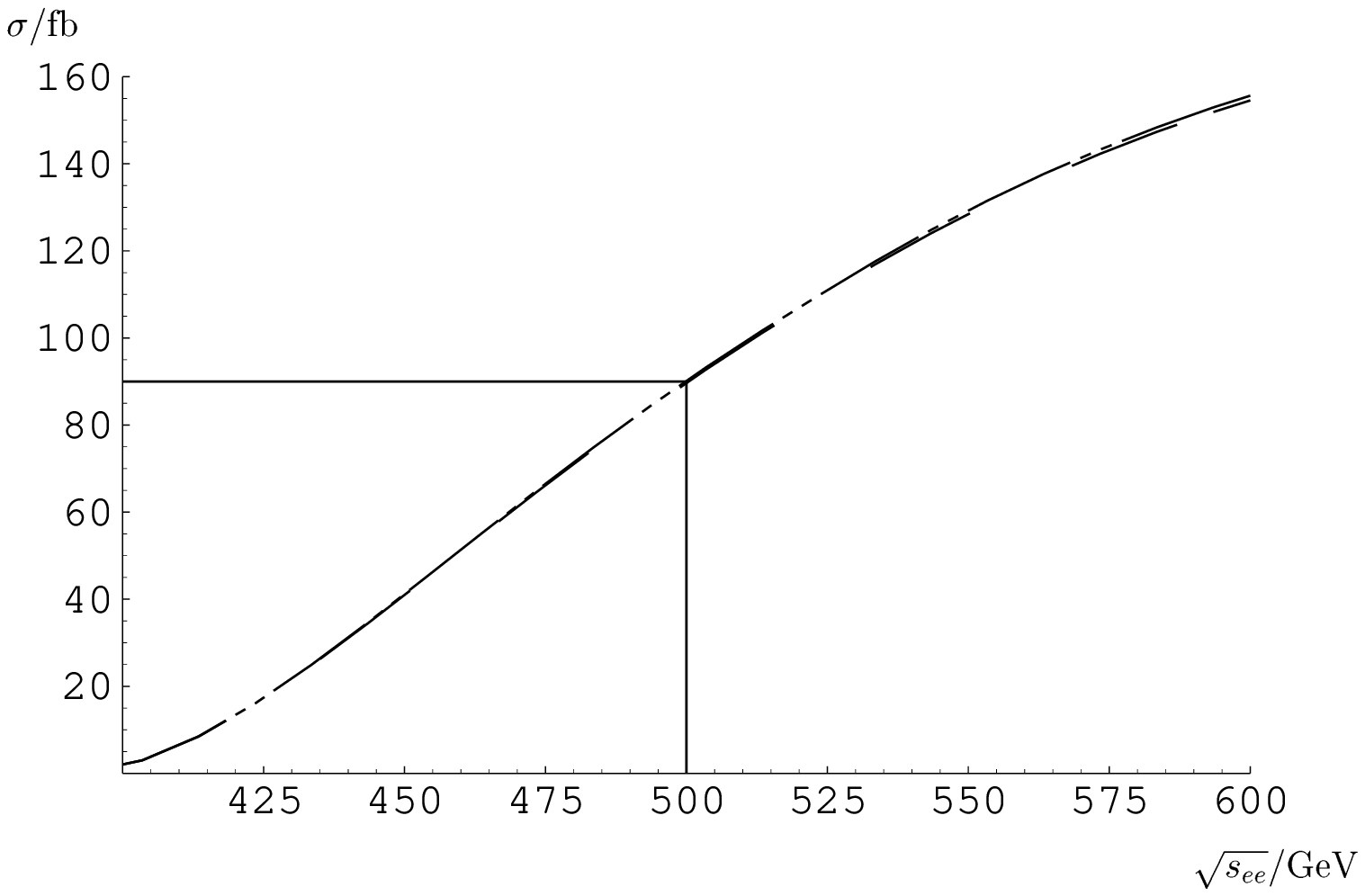}}
\put(2.4,-6.4){\includegraphics{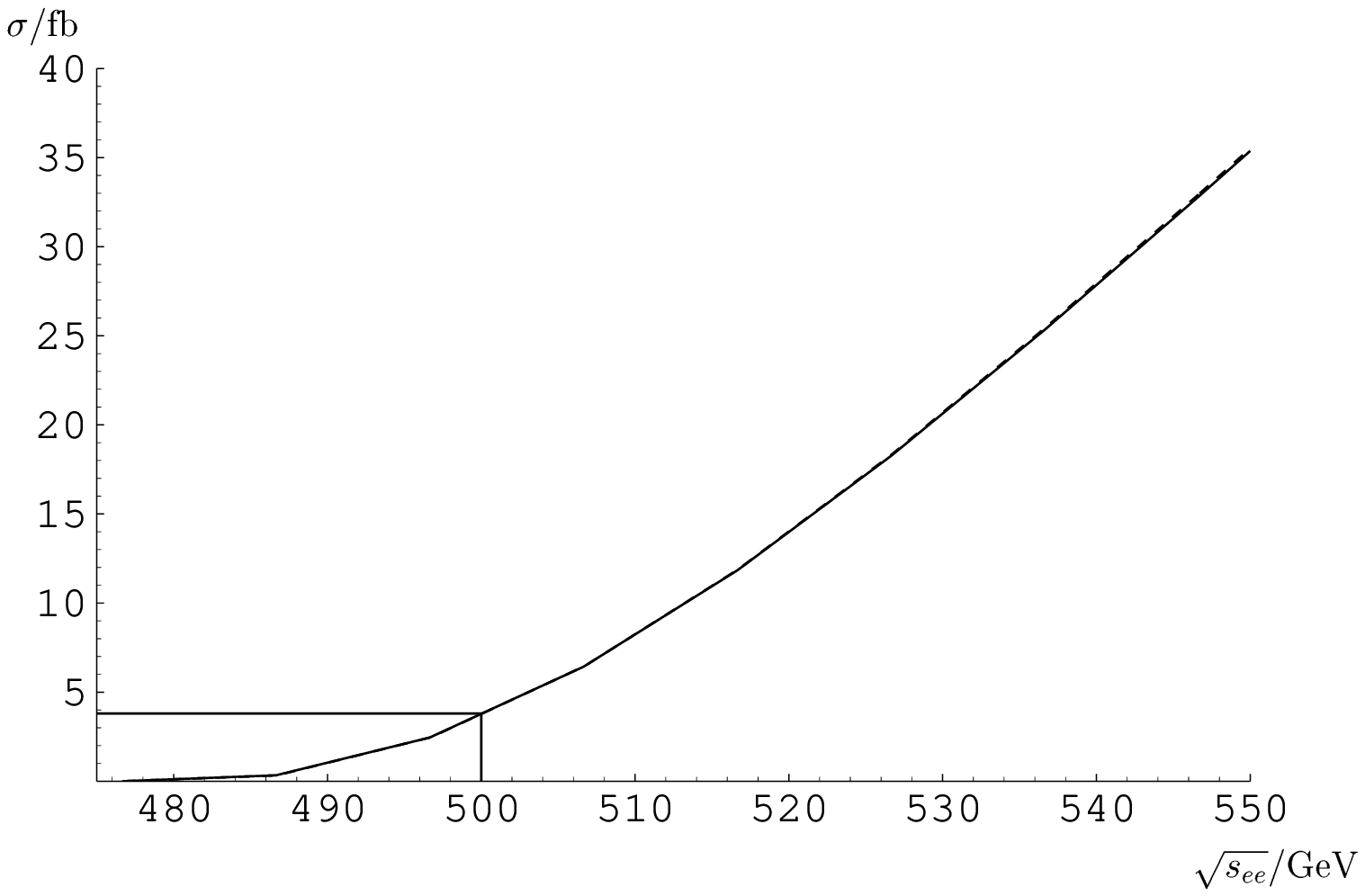}}
\end{picture}
\caption{Total cross sections in scenario I (full line), 
scenario II (dotted line), 
 scenario III (dashed line) and 
 scenario IV (dashed-dotted line): (a)
$\sigma\left(e^-e^+\rightarrow\tilde{\chi}_1^0\tilde{\chi}_2^0\right)$ 
with longitudinal polarisation 
$P_{e^-}=80\%$ and 
$P_{e^+}=-60\%$;
(b) 
$\sigma\left(e^-\gamma\rightarrow \tilde{\chi}_1^0\tilde{e}_{R}\rightarrow e^-\tilde{\chi}_1^0\tilde{\chi}_1^0\right)$ with polarisations $P_e=80\%$,
$\lambda_e=-80\%$, $\lambda_L=100\%$;(c) 
$\sigma\left(e^-\gamma\rightarrow \tilde{\chi}_2^0\tilde{e}_{R}\rightarrow e^-\tilde{\chi}_2^0\tilde{\chi}_1^0\right)$ with same polarisations as in (b).}
\end{figure}

\subsection{If $\tilde{\chi}_2^0$ is singlino-like ...}

... we expect for $e^+e^-\rightarrow \tilde{\chi}_1^0\tilde{\chi}_2^0$ 
a production cross section $\sigma=1.1$ fb in scenario III and 
$\sigma=0.31$ fb in scenario IV at $\sqrt{s_{ee}}=500$ GeV. 
If one assumes these 
cross section to be detectable one again can use 
$e\gamma$ scattering to decide which neutralino is singlino dominated. 
The detection of the process
$e^-\gamma\rightarrow \tilde{\chi}_1^0\tilde{e}_{R}\rightarrow e^-\tilde{\chi}_1^0\tilde{\chi}_1^0$ would prove the singlino character of $\tilde{\chi}_2^0$. 
In scenario III and IV one obtains $\sigma=90$ fb for $\sqrt{s_{ee}}=500$ GeV 
(fig.~2b).  

If, however,  the cross section for associated neutralino production in $e^+e^-$ annihilation
can not be observed, the process $e^-\gamma\rightarrow \tilde{\chi}_1^0\tilde{e}_{R}\rightarrow e^-\tilde{\chi}_1^0\tilde{\chi}_1^0$ would be the first
process for direct production of neutralinos. But this process alone does not 
aalow to discriminate between the MSSM and the NMSSM.

\section{Conclusion}

\begin{itemize}

\item For a singlino dominated $\tilde{\chi}_1^0$ the associated production of 
selectrons and the LSP in $e\gamma$-scattering
will be strongly suppressed.

\item $e\gamma\rightarrow\tilde{\chi}_{1/2}^0\tilde{e}_{L/R}$ could be the 
first process to identify
neutralinos and to determine the underlying supersymmetric model.

\item If neutralino pair production 
$e^-e^+\rightarrow\tilde{\chi}_1^0\tilde{\chi}_2^0$ gives evidence for
a neutralino with singlino character,
the $e\gamma$ process can be used to decide, whether
$\tilde{\chi}_1^0$ or $\tilde{\chi}_2^0$ is singlino dominated.
\end{itemize}

\end{document}